\documentstyle[aps,epsfig,floats,amssymb]{revtex}
\draft
\begin{document}
%two columns:
\twocolumn[\hsize\textwidth\columnwidth\hsize\csname
@twocolumnfalse\endcsname

\title{Conformal fixed point, Cosmological Constant and Quintessence }

\author{Christof Wetterich}

\address{
Institut f{\"u}r Theoretische Physik,
Philosophenweg 16, 69120 Heidelberg, Germany}

\maketitle

\begin{abstract}
We connect a possible solution for the ``cosmological constant problem'' to the existence
of a (postulated) conformal fixed point in a fundamental theory. The resulting cosmology
leads to quintessence, where the present acceleration of the expansion of the universe
is linked to a crossover in the flow of coupling constants.

\end{abstract}
\pacs{PACS numbers: 98.80. Cq, 95.35.+d  \hfill HD-THEP-02-37}

%two columns:
 ]

%%%%%%%%%%%%%%%%%%%%%%%%%%%%%%%%%%%%%%%%%%%%%%%%%%%%%%%%%%%%%%%%%%%%%%%%

Once upon a time gravity was a strong force, with Newton's constant $G^{(i)}_{eff}=
10^{110}m^3kg^{-1}s^{-2}$
or effective Planck mass $\bar{M}^{(i)}=2\cdot10^{-33}eV$, and typical particle masses
$\approx\bar{M}^{(i)}$. The mysterious homogeneous dark energy
in the universe (''cosmological constant'') started out with a similar characteristic
magnitude $V^{(i)}\approx (\bar{M}^{(i)})^4$. Over the ages of the history of the
universe the Planck mass increased and reaches today the value
$\bar{M}^{(0)}=2.44\cdot 10^{18}~GeV$. In the later stages of cosmology the mass ratios
$M_W/\bar{M}$ and $m_p/\bar{M}$ for the Fermi scale and the proton mass have been
approximately time independent. The growth rate of the dark energy was slower, however,
such that today $V^{(0)}=(2.2\cdot 10^{-3}eV)^4$, explaining one of the smallest numbers
observed in nature, $V^{(0)}/\bar{M}^4=6.5\cdot 10^{-121}$. In the present epoch the pace of change
of the fundamental mass scales slows down considerably, resulting in an accelerated
expansion of the universe. This tale of the cosmological history may seem somewhat
weird at first sight - we will argue here that it could naturally be associated
to the properties of a (postulated) conformal fixed point of a (still unknown) theory
unifying all interactions.

Our basic assumption states that in a fundamental theory of all interactions (FT)
all mass scales of particle physics are determined by a field $\chi$ rather than by a
fundamental constant. This is common in grand unified, higher dimensional or superstring
theories.
Typically, $\chi$ is associated to a scale of transition such that for momenta
$p^2\gg \chi^2$ all the modes of the FT are important - for example, the FT may be
formulated in more than four spacetime dimensions - whereas for $p^2\ll\chi^2$ an effective
description in terms of a four dimensional quantum field theory becomes valid. From the
FT point of view the field $\chi$ plays the role of an effective infrared scale. Within
the four dimensional description that we adopt here $\chi$ is a scalar field and may
therefore evolve over cosmological time scales. In particular, the effective Planck mass
is proportional to $\chi$. If $\chi$ changes with time one is led to cosmologies with a
variable Planck mass \cite{CW1}.

Dilatation or scale transformations correspond to a multiplicative rescaling
$\chi\rightarrow c\chi$, with constant $c$ and appropriate scaling of the metric and other
fields.
A nonvanishing cosmological value $\chi(t)$ ``spontaneously breaks'' dilatation and conformal
 symmetries and induces    masses for most particles.
If dilatation symmetry were an exact symmetry of the effective action
the value of $\chi$ would not be an
observable quantity. However, in quantum theories it is common that dilatation
symmetry is violated  \cite{PSW}, \cite{CW2}. by the
effects of fluctuations, resulting in ``running'' dimensionless couplings depending on
$\chi$. We assume that by dimensional transmutation this introduces an
intrinsic scale $m$ in the effective potential for the cosmon field $\chi$, in anlogy
to the characteristic scale of strong interactions, $\Lambda_{QCD}$.

As an example, we realize these ideas in an effective model for gravity and the cosmon
field $\chi$, characterized by an effective action $S$ after ``integrating out'' the other
fields and all quantum fluctuations
\begin{eqnarray}\label{1}
S&=&\int d^4x\sqrt{g}\Big\{-\frac{1}{2}\chi^2R\nonumber\\
&&+\frac{1}{2}\Big(\delta\left(\frac{\chi}{m}\right)-6\Big)\partial^
{\mu}\chi\partial_{\mu}\chi+V(\chi)\Big\}.
\end{eqnarray}
Here we consider the case where for $m\rightarrow 0$ the effective potential has a flat
direction, and assume that for large $\chi$ the leading manifestation of the dilatation
anomaly for $V$ results in a mass term
\begin{equation}\label{2}
V=m^2\chi^2.
\end{equation}
In the region of large $\chi\gg m$ all interactions are derivative interactions.
The dimesionless coupling $\delta>0$ governs the cosmon kinetic term. The additive
constant is chosen such that the model exhibits an exact local conformal symmetry
$g_{\mu\nu}\rightarrow c^{-2}(x)g_{\mu\nu},\chi\rightarrow c(x)\chi$
for $\delta=0~,~m=0$. In our normalization $\chi$ corresponds to the effective
reduced Planck mass, $\bar{M}=(8\pi G_{eff})^{-1/2}$.

It is straightforward to solve the field equations for this model \cite{CW1}\cite{CW3}
for a homogeneous and isotropic metric and cosmon field. One finds that the cosmon
field increases for large time! This is due to its coupling to gravity and contradicts
the too naive expectation that the cosmon should approach the potential minimum at
$\chi=0$. The increase of $\chi$ has a striking consequence for the fate of the
homogeneous dark energy. Indeed, only mass ratios are physically observable \cite{CW1}.
For the effective ``cosmological constant'' $V/\bar{M}^4$ we find that $V^{1/4}$
increases less rapidly than $\bar{M}=\chi$. Therefore the dark energy vanishes
asymptotically if $\chi$ increases with time, i. e. $V/\bar{M}^4= m^2/\chi^2
\rightarrow 0$. This is the basic ingredient for our explanation \cite{CW2} why the
``cosmological constant'' vanishes asymptotically and why dark energy has attained an
extremely small value today as a consequence of the enormous age of our universe.

Before we can proceed to a quantitative discussion of cosmology we need to specify
$\delta(\chi/m)$. For large $\chi\gg m$ simple dimensional arguments tell us that the
$\chi$-dependence can be written in terms of a ``renormalization group equation''
$\partial\delta/\partial\ln\chi=\beta_\delta(\delta)$.
A computation of $\beta_{\delta}$ would need the knowledge of the FT since the dominant
contributions arise from modes with $p^2\approx\chi^2$. We only know that $\beta_{\delta}$
should have a zero for $\delta=0$, since $\delta=0$ corresponds to an enhanced
(conformal) symmetry and separates a stable model for $\delta \geq 0$ from an unacceptable
unstable model for $\delta<0$. By continuity, for small enough $\delta$ the $\beta$-
function is also small and $\delta$ increases only slowly with $\chi$ (assuming
$\beta_{\delta}\geq 0$). For $\chi$ varying over many orders of
magnitude during the cosmological evolution we may nevertheless be confronted with a
situation where $\delta$ has grown large at some critical value $\chi_c$. At this scale
we expect a {\em crossover} from the vicinity of the conformal fixed point at
$\delta=0$ to an unknown behavior for large $\delta$. The crossover scale $\chi_c$ can
play an important role in cosmology. In particular, we will discuss a scenario where $\chi$
reaches $\chi_c$ in the present epoch, triggering an accelerated expansion of the
universe \cite{AE}.

As a simple example we take
\begin{equation}\label{4}
\frac{\partial\delta}{\partial \ln\chi}
=\beta_{\delta}=E\delta^2~,~\delta=
\frac{1}{E\ln(\chi_c/\chi)}~,
\end{equation}
where $\chi_c$ depends on $E$ and the ``initial value'' $\delta_i=\delta(\chi=m)$.
For small $E\delta_i$ the separation between the crossover scale $\chi_c$ and the
intrinsic scale $m$ becomes exponentially large
\begin{equation}
\frac{\chi_c}{m}=\exp\left(
\frac{1}{E\delta_i}\right)~,
\end{equation}
in close analogy to the inverse ratio between the strong interaction scale
$\Lambda_{QCD}$ and the unification scale. In particular, if $\chi_c$ is associated
to the present value $\bar{M}^{(0)}=\bar{M}_p=2.44\cdot 10^{18}GeV$ and
$E\delta_i\approx1/138$ we obtain a
present value for the potential part of the dark energy
\begin{eqnarray}
V^{(0)}&=&m^2\chi^2_c=\exp\left(-
\frac{2}{E\delta_i}\right)\bar{M}^4_p\nonumber\\
&=&6.56\cdot 10^{-121}\bar{M}^4_p=(2.2\cdot 10^{-3}eV)^4.
\end{eqnarray}

We emphasize that the cosmological evolution for this class of models is independent
of the initial conditions since the late time behavior is governed by a cosmic
attractor solution \cite{CW2}\cite{Rat}\cite{Lid}. Generically, the ratio $\Omega_h$ between the
homogeneous dark energy density and the critical energy density stays small as long as
$\delta$ is small, adjusting itself to a dominant radiation or matter component,
$\Omega_h\approx\delta$ or $\Omega_h\approx\frac{3}{4}\delta$, respectively. This  behavior only
changes once $\chi$ reaches the crossover scale $\chi_c$, and for $E\delta_i\approx 1/138$
this happens precisely at the present epoch. Then the universe switches to a regime
where dark energy dominates.

To be more quantitative we select\footnote{We have divided $\beta_\delta$
in eq.(\ref{4}) by $(1+0.05\delta)$ in order
to keep $\delta$ finite for all $\chi$, without affecting the behavior for small
and intermediate $\delta$.} $E=5~,~\delta_i=1.444\cdot 10^{-3}$. We can now compute
the characteristic
quantities like the amount of dark energy today, $\Omega^{(0)}_h=0.7$, or the
equation of state
$w_h=p_h/\rho_h$ at the present time, $w^{(0)}_h=-0.93$. They are
compatible with the supernovae observations \cite{AE} and the age of the universe $t^{(0)}
=13.7\cdot 10^9yr$. For a discussion \cite{Doran} of the spectrum of the CMB-unisotropies
we need,
in addition, the value of $\Omega^{(ls)}_h=0.019$ at the time of last scattering and an
averaged equation of state $\bar{w}$ which determine the position of the third peak
in angular momentum space as $l_3=795$ (for
$h=0.66)$. Structure formation is slowed down by the early presence of dark energy \cite{Fer}.
It depends on an average of $\Omega_h$ over the time of structure formation,
$\Omega^{(sf)}_h=0.037$ \cite{Schwindt}. In our case the
cold dark matter density fluctuations
are reduced by a factor $\sigma_8/\sigma_8(\Lambda)=0.7$ as compared to a
model with a cosmological constant and the same amount of dark energy today. We conclude
that our simple model is compatible with the present observations. We observe interesting
differences as compared to models with a cosmological constant. They are subject to future
observational tests.

Several comments are in order: (i) Cosmology is most easily discussed after a Weyl scaling
$g_{\mu\nu}\rightarrow(\bar{M}_p/\chi)^2g_{\mu\nu}$ and a redefinition of the cosmon field
$\varphi/\bar{M}_p=\ln\big(\chi^4/V(\chi)\big)=2\ln(\chi/m)$, such that the coefficient in front of the
curvature scalar $R$ becomes constant and $\varphi$ is directly related to the value of the
potential\\
\begin{eqnarray}
S&=&\int d^4x\sqrt{g}\Big\{-\frac{1}{2}\bar{M}^2_pR\nonumber\\
&&+\frac{1}{2}k^2
(\varphi)\partial^{\mu}\varphi\partial_{\mu}\varphi
+\bar{M}^4_p\exp\left(-\frac{\varphi}{\bar{M}_p}\right)\Big\}.
\end{eqnarray}
The details of the model are now encoded
in the nontrivial kinetic term \cite{Hebecker} $k^2(\varphi)=\delta/4$. As
a general feature, the
motion of the cosmon slows down once $k^2$ becomes large, in our case for $\varphi$ near
$\varphi_c=2\bar{M}_p/(\delta_iE)$. Then the dominance of the
potential $V$ over the kinetic energy $T$ leads to a negative equation of state
$w_h=(T-V)/(T+V)$ and to an acceleration of the universe \cite{Stei}.

ii) The qualitative features of our proposal hold for a much more general
class of cosmon potentials $V$. For example, adding to eq. (\ref{2}) a ``bare cosmological
constant'' $\gamma m^4$ becomes completely irrelevant for large $\chi/m$. Only the behavior
of $V$ for large $\chi$ matters.
The scenario of an asymptotically vanishing dark energy holds provided that
$V$ increases less rapidly than $\chi^4$ and $\delta$ remains finite for $\chi<\bar{M}_p$.
This applies,
in particular, to an asymptotic behavior
$V=\lambda(\chi/m)\chi^4$ with a dimensionless coupling $\lambda$ obeying
the renormalization group equation
\begin{equation}\label{9}
\frac{\partial\lambda}{\partial\ln\chi}=-A\lambda~,~A>0.
\end{equation}

iii) Details of cosmology depend on $\beta_{\delta}$. For $\beta_{\delta}=0$ one recovers
``exponential quintessence'' \cite{CW2}, whereas for $\beta_{\delta}=D\delta,~D$ constant, one
finds ``inverse power law quintessence'' \cite{Rat} with power $\alpha=2A/D$.
We have studied other models with crossover behavior, e. g.:
$\beta_{\delta}=D\delta+E\delta^2$. For $D>0$ the required value of $\delta_i$ decreases
and early quintessence (e. g. $\Omega^{(ls)}_h~,~\Omega^{(sf)}_h)$ becomes less important.
The precise flow for very large $\delta$, i. e. $\delta$ remaining finite for all
$\chi$, plays only a minor role for presently observable cosmology provided $\delta$
grows sufficiently large in the present epoch.

iv) We can extend our description to matter fields and radiation. As an example we
consider the Higgs doublet $H$, a fermion field $\psi$ and the gluons characterized by
their field strength $F_{\mu\nu}$. Within our assumption this adds to eq. (\ref{1}) a
term
\begin{eqnarray}
S_M=\int d^4x\sqrt{g}\Big\{(\lambda_H/2)(H^{\dagger}H-\beta^2\chi^2)^2\nonumber\\
+(h\bar{\psi}_LH\psi_R+h.c.)+\frac{Z_F}{4}F^{\mu\nu}F_{\mu\nu}\Big\}.
\end{eqnarray}
Our previous description has neglected the possible $\chi$-dependence of the dimensionless
couplings $\lambda_H,\beta,h,Z_F$ such that after the Weyl scaling $\varphi$ decouples
completely from matter and radiation. In this case the Higgs doublet reaches its $\chi$-
dependent minimum $|H|^2=\beta^2\chi^2$ early in cosmology (after the electroweak phase
transition). Similarly, for a fixed value of the running gauge coupling at some grand
unified scale $M_X~,~\alpha_S(M_X)\approx 1/40$, we find $\Lambda_{QCD}\sim\chi$ if
$M_X\sim\chi$. In this approximation all ratios of particle masses become independent
of $\chi$ and do not vary with cosmological time.

We note the appearance of two different
types of characteristic masses for the excitations. The excitation along the ``vacuum
direction'' corresponds to a simultaneous change of {\em all} mass scales (along the direction
$|H|=\beta\chi)$. Its mass is given by the small intrinsic mass $m$. On the other hand, the
excitations perpendicular to the ``vacuum direction'' correspond to a variation of mass
ratios and have a characteristic mass $\sim\chi$. In our example, the intrinsic mass $m$
is many orders of magnitude smaller than the (present) mass of the Higgs boson $M^2_H
=\lambda_H\beta^2\chi^2$. This resembles the  spontaneous breaking of
some unknown  global symmetry where  a small mass $m$ for the pseudo-Goldstone boson
$\chi$ is induced by an anomaly.

v) We do not expect the dimensionless couplings $\lambda_H,\beta,h,Z_F$ to be precisely
independent of $\chi/m$. Then the cosmological variation of $\chi/m$ will induce a time
dependence of the fundamental parameters. Severe bounds \cite{Uz} restrict
\cite{CW1}\cite{CWVC} this dependence for the dimensionless couplings of the known
fields. More freedom is left for a coupling of the cosmon $\varphi$ to dark matter-
sizeable couplings would influence the cosmology \cite{CW3}\cite{Amendola}. Very close
to the big bang, for $\chi\approx m$, the dependence of all couplings on
$\chi/m$ may have been strong.

vi) In a grand unified theory the renormalized strong gauge coupling or the fine structure
constant $\alpha_{em}$ depend on the value of the gauge coupling $g_X=g(M_X)$ at the
unification scale $M_X$ where $g^2_X(\chi)\sim Z^{-1}_F(\chi)$. We neglect
 here for simplicity the $\chi$-dependence of
$M_X~,~B_X=-\partial\ln(M_X/\chi)/\partial\ln\chi\approx 0$, and concentrate on the
case where for $\chi\rightarrow\infty$ the running of $g_X$ is governed by a fixed point
$g^2_*/4\pi\approx 1/40$
\begin{eqnarray}\label{11}
\frac{\partial g^2_X}{\partial\ln\chi}=\beta_{g^2}&=&
b_2g^2_X-b_4g^4_X~,~\nonumber\\
b_2&=&b_4g^2_*>0.
\end{eqnarray}
It is interesting to associate $m$ with the nonperturbative scale where
$g_X(\chi\rightarrow m)\rightarrow\infty~,$
\begin{equation}
g^2_X=g^2_*\left[1-\left(\frac{\chi}{m}\right)^{-b_2}\right]^{-1}.
\end{equation}
The present relative variations of the gauge couplings are then determined by
\begin{eqnarray}
\eta_F=-\frac{\beta_{g^2}}{g^2_X}\approx b_2\left(\frac{\chi}{m}\right)^{-b_2}
=\exp\left(-\frac{b_2\varphi}{2\bar{M}_p}\right).
\end{eqnarray}
For sufficiently large $b_2$, say $b_2>0.2$, the time variation of $\Lambda_{QCD}$ or
$\alpha_{em}$ is much too small to be accessible for present observations \cite{CWVC}. More
generally, we conclude that a fixed point which is approached sufficiently fast for
$\chi\rightarrow\infty$ could give a very simple explanation why the cosmological
time variation of fundamental couplings is small. On the other hand, the substantial
variation of $\delta$ at the present cosmological epoch may have a small influence on the
precise location of the fixed point $g_*(\delta)$.
A small increase of $g_*(\delta)$ can lead to a
time variation of $\alpha_{em}$ in the range inferred from the observation of quasar
absorption lines \cite{QSO}, corresponding to $\eta_F=-4\cdot10^{-7}$. Recent
investigations indeed show \cite{RC} that a small $\delta$-dependence of $\beta_{g^2}$
makes the QSO observation compatible with all present bounds from archeo-nuclear physics
and tests of the equivalence principle, provided that the crossover is sufficiently rapid
(e.g. $E=8$).

vii) The effect of the quantum fluctuations is encoded in the
$\beta$-functions  (\ref{9}), (\ref{11}). In our setting this concerns mainly
small deviations from a
conformal fixed point at $\delta=0~,~\lambda=0~,~g^2=g^2_*$. We emphasize that the
existence of a fixed point at $\lambda=0$ seems plausible since it separates again
a stable $(\lambda>0)$ from an unstable $(\lambda<0$, unbounded potential) situation.

The conformal fixed point has $g^2$ (and $\lambda$) as relevant coupling for
$\chi\rightarrow 0$ whereas for $\chi\rightarrow\infty$ the relevant coupling corresponds
to $\delta$. The flow of the couplings is therefore neither stable towards the infrared
nor towards the ultraviolet. The scale $m$ marks the first infrared scale where couplings
grow large (e. g. the gauge coupling) whereas $\chi_c$ corresponds to the instability
in the ultraviolet. A huge ratio $\chi_c/m$ occurs whenever the trajectory of the
flow passes sufficiently close to the fixed point.

A crucial question concerns the running of $\lambda$ for large $\chi$  which can depend on
the various dimensionless couplings
$\partial\lambda/\partial\ln\chi=\beta_\lambda(\lambda,\delta,g^2,h,\lambda_H,\dots).$
Near a fixed point with only one relevant (marginal) coupling $(\delta)$ all couplings
follow critical trajectories which may be parameterized by
$\delta(\lambda),g^2(\lambda),h(\lambda)$ etc. Inserting these functions into
$\beta_\lambda$ yields an expansion for small $\lambda,\beta_\lambda=c_\lambda-A\lambda+\dots$.
Only for $c_\lambda=0$ the fixed point occurs for $\lambda_*=0$ and only in this case the
effective cosmological constant vanishes asymptotically. Our assumption of a flat
direction in the effective cosmon potential is equivalent to $c_\lambda=0$.
 The individual contribution of a particle with mass
$M_j=\gamma_j\chi$ is $c_\lambda\sim\gamma^4_j$. Even though $\gamma_j$ is a tiny coupling
for standard model particles the resulting cosmological constant would come out much
too large $(\sim M^4_j)$, reflecting the ``naturalness'' or ``fine tuning'' problem.

It is obvious, however, that $c_\lambda$ will be dominated by the unknown particles with mass
around $\chi$. From the point of view of the low energy theory the value of $c_\lambda$ is
an {\it ultraviolet} problem. We see no way to make statements about the location of
$\lambda_*$ from our knowledge of the low energy theory \footnote{The situation is very
different if the location of a fixed point is dominated by the infrared modes! Actually,
an attempt to characterize the location of an ultraviolet fixed point by the couplings
of the infrared modes would be similarly misleading as the characterization of an infrared
fixed point by properties of the ultraviolet modes.}.
This argument is strengthened if $\gamma_j$ depends on $\chi$ as in our setting. In this
case even the contribution of the low mass particles to $c_\lambda$ is not uniquely
dominated by momenta $q^2\sim M^2_j$ - it also involves ``ultraviolet momenta''
$q^2\approx\chi^2$. We think that in view of this situation our conjecture
that $V(\chi)$ rises for large $\chi$ less rapidly than $\chi^4$ (coresponding to
$\lambda_*=0)$ seems to be a reasonable possibility \footnote{Our setting circumvents
an argument \cite{QCO} that time varying fundamental constants require the tuning of a
whole function. Only $c_\lambda=0,A>0$ is required.}. In this respect it is crucial that the
late time behavior explores the ultraviolet rather than the infrared!

We conclude that in a fundamental theory the presence
of a conformal fixed point with a flat direction, together with
the flow of small deviations from the fixed point proposed in this note $(A>0)$, would lead
to a natural explanation why the cosmological constant vanishes for asymptotic time.
After Weyl scaling, no additive constant hinders the asymptotic approach of the cosmon
potential to zero, $V(\varphi\rightarrow\infty)\rightarrow 0$.
This would solve the ``cosmological constant problem'' \cite{Wei}.
In our crossover scenario
the past evolution of the universe is characterized by a small and slowly varying fraction
of dark energy which adapts to a dominant radiation (matter) component
$\Omega_h\approx\delta\left(\frac{3}{4}\delta\right)$. The future of the universe depends
crucially on the unknown properties of the flow of $\delta(\chi)$ in the region of the large
$\delta$. The present epoch witnesses a crossover from small to large $\delta$, resulting
in an accelerated expansion. In a FT the crossover scale $\chi_c/m$ should be computable,
just as the values of mass ratios or dimensionless couplings in particle physics. It is
therefore not excluded that some of the ``cosmic coincidences'' (relations between the
present value of the Hubble parameter $H_0$ and particle properties) could find an
explanation in this direction. For the potential (\ref{2}) the present value of $H$ is
given by the mass $m$ characterizing the dilatation anomaly
\begin{equation}
H^2_0=\frac{2\Omega^{(0)}_h}{3(1-w^{(0)}_h)}m^2.
\end{equation}

The time variation of the dark energy could be detected by cosmological observations in the
near feature, and an establishment of a time variation of fundamental couplings would be a
striking argument in favor of our proposal.

\end{document}